\documentclass[12pt, preprint]{emulateapj}
\usepackage{hyperref}
\hypersetup{colorlinks,
 citecolor=blue,
 linkcolor=blue}
\usepackage{soul}
\usepackage{natbib}

\slugcomment{Draft Version \today}
\shorttitle{Stellar populations of UCDs vs.\ dE nuclei}
\shortauthors{Paudel et al.}
\begin{document}
\title{Nuclei of early-type dwarf galaxies: are they progenitors of
  UCDs?\altaffilmark{1}}

\author{Sanjaya Paudel\altaffilmark{2,3}, Thorsten Lisker\altaffilmark{2},
 Joachim Janz\altaffilmark{4,2,5}
  }
\affil{
$^2$ Astronomisches Rechen-Institut, Zentrum f\"ur Astronomie der
  Universit\"at Heidelberg, \\M\"onchhofstra\ss e 12-14, 69120 Heidelberg, Germany\\
$^4$ Division of Astronomy, Department of Physical Sciences,
  University of Oulu, P.O.\ Box 3000, FIN-90014 Oulu, Finland\\
}

\altaffiltext{1}{Based on observations collected at the European 
    Organisation for Astronomical Research in the Southern Hemisphere,
    Chile (programs 078.B-0178 and 085.B-0971)}
\altaffiltext{3}{Email: sjy@x-astro.net}
\altaffiltext{5}{Fellow of the Gottlieb Daimler and Karl Benz Foundation}
\begin{abstract}
To address the question of whether the so-called ultra compact dwarf
galaxies (UCDs) are the remnant nuclei of destroyed early-type dwarf galaxies
(dEs), we analyze the stellar population parameters of the nuclei of 34
Virgo dEs, as well as ten 
Virgo UCDs, including one that we discovered and which we report on here. Based on absorption line strength (Lick
index) measurements, we find that nuclei of
Virgo dEs have younger stellar population ages than UCDs, with 
averages of 5 Gyr and $>$10 Gyr, respectively. In addition
to this, the metallicity also differs: dE nuclei are on average more metal-rich
than UCDs. On the other hand, comparing the
stellar population parameters at the same local galaxy density, with
UCDs being located in the high density cluster regions, we do not find
any difference in the stellar populations of dE nuclei and UCDs. In those regions, the dE nuclei are as
old and as metal poor as UCDs. This evidence suggests that the Virgo
UCDs may have formed through the stripping of dE nuclei.

\end{abstract}


\section{Introduction}

Since the discovery of ultra-compact dwarf galaxies (UCDs; \citealt{Hilker99};
\citealt{Phillipps01}), it is still a complicated puzzle in extragalactic
astronomy how such compact and luminous objects may have
formed. They are brighter and larger than globular clusters
(GCs) \citep{Mieske02} and much smaller than early-type dwarf
galaxies (dEs) in both size and luminosity. A number of studies
targeting various UCD samples in different galaxy clusters also
revealed the diverse nature of UCDs: Fornax UCDs are slightly redder
on average than Virgo UCDs
\citep{Evstigneeva08,Mieske08aa,Mieske06}. On the other hand, it is
still a matter of debate whether or not the UCDs contain dark matter
\citep{Drinkwater03,  Hasegan05, Hilker07}. This makes them very 
special objects to study in extragalactic astronomy, suggesting that
the presence of dark matter or not can be directly related to whether 
UCDs have a galactic origin or not.

Overall it has been already noted that Virgo UCDs contain fairly old
(age: $>$ 8-10 Gyr) and metal poor ($< -$0.5 dex) stellar populations
\citep[hereafter E07]{Evstigneeva07}. Therefore, it is also proposed
that they could be very luminous intra-cluster GCs
\citep{Mieske02}. Another popular formation
scenario is the threshing of nucleated dEs 
\citealt{Bassino94}. In this picture, UCDs are the remnants of galaxies that have been
significantly stripped in the cluster environment. Numerical
simulations \citep{Bekki03,Goerdt08} have generally confirmed that the remnant
nuclei resemble UCDs in their structural parameters.

Stellar population studies of dEs  provide evidence that the
nuclei have intermediate ages and moderately metal-enriched
stellar populations \citep{Koleva09, Chilingarian09}. In addition to this, 
since UCDs show slightly super solar [$\alpha$/Fe] abundances,
\citet{Evstigneeva07} argued that the stellar population properties
rather support the view that UCDs are luminous globular 
clusters than being nuclei of dEs.

In this letter, we present a stellar population analysis based on
absorption-line strengths (Lick indices, \citealt{Burstein84};
\citealt{Worthey94a}; \citealt{Trager98}) of a fairly large sample
of 34 nucleated dEs and 10 UCDs in the Virgo cluster. So far, studies comparing
stellar population parameters derived from spectra used rather low
numbers of objects. Moreover, the extraction of nuclear spectra has been made without subtracting
the underlying galactic light, which can still contribute significantly at the photometric
center of the dEs. We therefore apply a simple method to subtract most of
this light (see Section \ref{reduc}), thus expecting that our
measurements are representative for the stellar population properties
of the nuclei themselves.
Finally, we present the distributions of the stellar population parameters of
 dE nuclei and UCDs with respect to local galaxy density and to their
 luminosity, and we try to constrain possible formation scenarios of Virgo UCDs.

\section{The sample, observations  and data reduction} \label{reduc}
\begin{figure}
 \includegraphics[width=8cm]{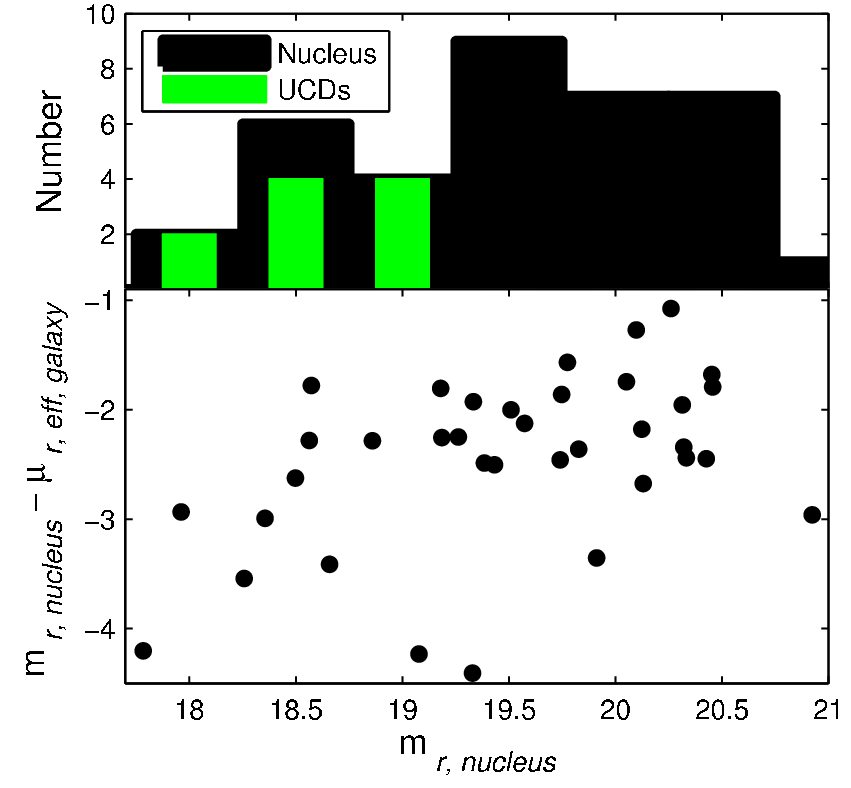}
 \caption{\emph{Top:} Number distribution of dE nuclei and UCDs in our
   sample. \emph{Bottom:} Our selection criteria for dE nuclei,
   limiting the nucleus magnitude $m_{r, {\rm nucleus}}$ to values
   brighter than 21 mag, and the ``nucleus strength'', i.e.\ the value of $m_{r, {\rm
       nucleus}}-\mu_{r, {\rm eff, galaxy}}$ to less than -1. Note
   that the nucleus sample is not complete within this parameter
   region, i.e.\ not all Virgo dE nuclei with these parameters have
   been observed.}
 \label{his}
\end{figure}

\subsection{Sample selection and a new Virgo UCD}
\label{scucd}

Our dE sample comprises 34 nucleated dEs in the Virgo cluster 
\cite[Virgo cluster catalog, VCC,][]{Binggeli85, Binggeli93},
selected to have a relatively high ``nucleus strength'' 
\cite[details of the sample see,][]{Paudel10}, which we
define as the difference between the nucleus magnitude
  and the host galaxy
effective surface brightness, $s_{r, \rm nucleus} = m_{r, {\rm
       nucleus}}-\mu_{r, {\rm eff, galaxy}}$, measured in SDSS $r$
(see below). Thereby, $\mu_{r, {\rm eff, galaxy}}$ is a measure for
the brightness of a unit area of the galaxy, determining the
``contrast'' between galaxy and nucleus (also see \citealt{Lisker07},
their Fig. 1).
 We
select nuclei with $s_{r, \rm nucleus}<-1$ and $m_{r, {\rm
    nucleus}}<21$ mag (see Fig.~\ref{his}).

Our UCD sample selection (see Table 1), is based on
\citet{Evstigneeva05} and \citet{Jones06}; our numbering follows the latter. Three of
the nine Virgo UCDs of Jones et al.\ were not included in the Lick
index study of Evstigneeva et al.\ (VUCD2, 8, and 9), so they were
selected by us as targets. Three further UCDs were selected, since they fell in the same
field-of-view as dE targets of our study. Due to the multi-slit
observations, they could be easily included.

We also targeted a new Virgo UCD candidate, which we now indeed
confirm as Virgo cluster member; it is named VUCD10 in
Table 1. It was identified through a simple multiparameter selection
procedure. From SDSS DR5 pipeline photometry, we obtained $ugriz$
magnitudes and colors for all nine Virgo UCDs in
\citet{Jones06}. When excluding VUCD7, which is clearly 
brighter than the others and appears to
be an extended object in the SDSS images (also see \citealt{Jones06}),
the {\it r}-band magnitudes (SDSS ``modelMag'' values) lie within 18.0
to 19.1 mag. Their Petrosian radii in $r$, again excluding 
VUCD7, are below 2.2 arcsec. Their {\it u-r} colors, when excluding the
much redder VUCD3, cover the range 1.8 to 2.4 mag (which includes
VUCD7). Their {\it i-z} colors, again excluding VUCD3, lie between
0.1 and 0.25 mag (which again
includes VUCD7). The right ascension and declination of all objects
except VUCD3 and VUCD7 ranges from 187.5$^\circ$ to 188.1$^\circ$ and 11.9$^\circ$ to
12.7$^\circ$, respectively. When querying the SDSS database for all objects
fulfilling the above criteria of magnitude, radius, color, and
position, 20 objects were identified that the SDSS classified as
stars, among them VUCD1, 2, and 5. The same query, but for objects
classified as galaxies, yielded only five objects: VUCD4, 6, 8, 9, and
the new VUCD10, which we therefore included in our target sample.

With its radial velocity of 2425 km/s that we now measured from its
spectrum, it is consistent with being a Virgo cluster member: in
velocity space, Virgo member galaxies in the central cluster region
reach velocities of 2600 km/s
(\citealt{Binggeli87,Binggeli93}).
We therefore consider VUCD10 
a new Virgo cluster UCD, and include it in the following analysis.

\subsection{Photometry}

To obtain nucleus and UCD magnitudes, we use SDSS DR5 $r$-band images to
which we applied our own sky subtraction procedure
\citep[see][]{Lisker07}. We also correct them for Galactic extinction,
following \citet{Schlegel98}, and use the flux calibration provided
directly by the SDSS. A Virgo distance modulus of $m-M=31.09$ mag
\citep{Mei07, Blakeslee09}, corresponding to $d=16.5$ Mpc and an SDSS pixel scale of
32 pc (80 pc/"), is adopted here and in the further analysis.

Nucleus magnitudes are
derived as follows. A S\'ersic fit to the radial
profile of the galaxy is done, measured with elliptical annuli and using only
the radial interval from 2" to 1/3 of the half-light radius, or
extending it if the interval subtends less than 2.5" (200 pc).
From this fit, a two-dimensional elliptical model image, taking
  into account the median SDSS PSF of 1.4" FWHM, is
created and subtracted from the original image, leaving only the
nucleus in the center. The nucleus magnitude is then measured by
circular aperture photometry with $r=2"$. We estimate the magnitude error to be 0.2
mag, which is consistent with the RMS scatter of a comparison to the {\it Hubble Space
  Telescope (HST)} ACS Virgo
cluster survey \citep{Cote06} for the 10 objects in
common.\footnote{Nucleus $r$ magnitudes were converted into $g$
    by adding the $g-r$ color value of the nucleated dE's center
    ($r\le2"$), and were then compared to ACS $g$ magnitudes, showing
    no systematic offset.}

The same aperture photometry
is done to obtain UCD magnitudes. To be comparable to the nucleus
measurements, this aperture is also applied to
VUCD7, even though it is known to be
extended, as mentioned previously. For the UCD magnitudes we estimate
an error of 0.1 mag.

\subsection{Spectroscopy}

The spectroscopic observations were carried out at ESO Very Large Telescope (VLT)
with FORS2. The 1" slit and V300 grism provide an
instrumental resolution $\sim$11 \AA{} FWHM.  Details of the
observational layout and data reduction are described in
\citet[hereafter P10]{Paudel10}. Our data was obtained in two
observing runs, using the same instrumental setup. 
The first set of dEs (26 objects, see P10) was observed in the semester 2007A,  and
the second set (8 dEs) in 2010A.
In total our sample covers the full range of local projected
galaxy density that is populated by nucleated dEs in Virgo, from less
than 10 to more than 100 galaxies per square degree. The
UCDs were observed in 2007A together with the first set of dEs, using
the same instrumental configuration. We provide the basic
properties such as position and radial velocity in Table \ref{ucds}
and the detailed UCD sample selection criteria in Section \ref{scucd}.

 While the nucleus typically contributes the majority of light
at the photometric center of the
dEs, still a considerable amount of underlying galactic light of the
host galaxy could alter the actual properties of the nuclei. Therefore,
we attempt to subtract the galactic light from the nuclear
spectra. This is done by inward extrapolation of the exponentially-fitted light profile of the host
galaxy along the slit, yielding the amount of galaxy light
contained in the nucleus extraction aperture of 0".75 (central 3
pixels). The galaxy spectrum is extracted in a radial interval from 3"
to 8", scaled accordingly, and subtracted from the centrally extracted
spectrum. Further details of this process are given in Paudel et
al.\ (2010b, submitted to MNRAS). While certainly not being perfect, we expect that this approach
removes most of the galactic light contamination and thus provides
spectra that are fairly well representative for the stellar population
properties of dE nuclei.

\begin{table*}
 \centering
  \caption{The UCD sample. In the first four columns, we provide
    UCD number, position in the sky (RA, Dec),
    and radial velocity. In the last column, we indicate whether UCDs
    are in
    common with the studies of E07 or \citet{Firth09}.  }
  \label{ucds}
 \footnotesize
\begin{tabular}{llllllllll}
\hline
UCDs	&	RA	&	Dec	&	RV		& Age & [Z/H] & [$\alpha$/Fe] & m$_{r}$ &	Remark	\\
name	&	(h:m:s) 	&	($^{o}$ :' :")	&	(kms)	& Gyr & dex & dex & mag	&		\\
\hline

VUCD1	&	12:30:07.61	&	12:36:31.10	&	1227.8	$\pm$	1.7	&  11.9$^{  +1.8}_{ -2.0}$ & -0.90 $\pm$ 0.05  &  0.38 $\pm$ 0.09   & 18.6  &	E07  \\   
VUCD2	&	12:30:48.24	&	12:35:11.10	&	911	$\pm$	6	&   5.9$^{  +1.7}_{ -0.5}$ & -0.77 $\pm$ 0.09  & -0.14 $\pm$ 0.08   & 18.6  &	F09  \\ 
VUCD3	&	12:30:57.40	&	12:25:44.8	&	710.6	$\pm$	3.5	&  15.0$^{  +0.1}_{ -1.3}$ &  0.13 $\pm$ 0.03  &  0.38 $\pm$ 0.03   & 18.1  &	E07  \\ 
VUCD4	&	12:31:04.51	&	11:56:36.8	&	919.7	$\pm$	1.7 	&  11.9$^{  +1.1}_{ -2.0}$ & -1.24 $\pm$ 0.05  & -0.00 $\pm$ 0.12   & 18.6  &	E07  \\ 
VUCD5 	&	12:31:11.90	&	12:41:01.20	&	1290	$\pm$	3	&  10.7$^{  +0.9}_{  0.1}$ & -0.40 $\pm$ 0.03  &  0.31 $\pm$ 0.02   & 18.5  &	E07  \\ 
VUCD6 	&	12:31:28.41	&	12:25:03.30	&	2101	$\pm$	4	&   8.3$^{  +6.7}_{ -1.3}$ & -1.03 $\pm$ 0.12  &  0.41 $\pm$ 0.09   & 18.7  &	E07  \\ 
VUCD7 	&	12:31:52.93	&	12:15:59.50	&	985	$\pm$	5	&  10.7$^{  +0.9}_{  0.1}$ & -0.54 $\pm$ 0.03  &  0.13 $\pm$ 0.04   & 18.0  &	E07  \\ 
VUCD8 	&	12:32:04.33	&	12:20:30.62	&	1531	$\pm$	19	&  11.6$^{  +1.0}_{ -0.9}$ & -1.24 $\pm$ 0.03  & -0.09 $\pm$ 0.09   & 19.1  &	F09  \\ 
VUCD9 	&	12:32:14.61	&	12:03:05.40	&	1325	$\pm$	5	&  11.6$^{  +1.0}_{  0.1}$ & -1.03 $\pm$ 0.06  &  0.07 $\pm$ 0.11   & 18.8  &	F09  \\ 
VUCD10	&	12:30:30.86	&	12:18:41.66	&	2425	$\pm$	103	&   9.8$^{  +2.8}_{ -0.8}$ & -0.89 $\pm$ 0.09  &  0.50 $\pm$ 0.04   & 19.1  &	New  \\

\hline
 \end{tabular}
\end{table*}

\section{Results}

\begin{figure}
 \includegraphics[width=8cm]{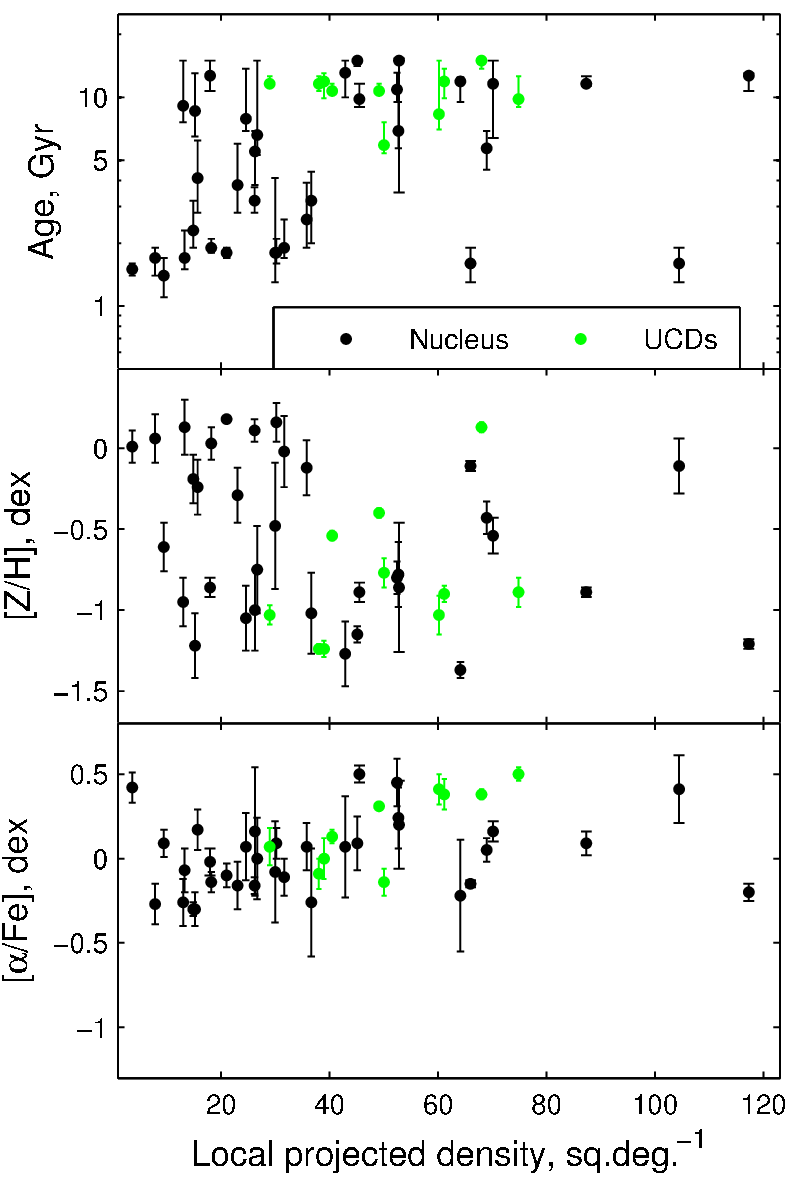}
 \caption{Age, metallicity, and [$\alpha$/Fe]-abundance versus local
   projected galaxy density are shown for dE nuclei (black) and UCDs
   (green).} 
 \label{env2}
\end{figure}

In order to derive stellar population parameters from the
measured line strengths, we translate our Lick index measurements into SSP-equivalent ages, metallicities, and
$\alpha$-element abundance ratios by comparing them to the model of
\citet{Thomas03} by $\chi^{2}-$minimization, following
\citet{Proctor02}.  We use nine well-measured indices (H$\delta_F$, H$\gamma_F$, Fe4383,
H$\beta$, Fe5015, Mgb, Fe5270, Fe5335 \& Fe5406) in case of the dE
nuclei. For the UCDs we use only H$\beta$,
Mgb, Fe5270 and Fe5335, because we only find these four indices
available in the literature for those three UCDs that we did not target
ourselves. Moreover, due to the slit mask placement, our observations
only cover the full range of indices (i.e., $\sim$4000 \AA{} to $\sim$5600\AA) for
three of the observed UCDs. For completeness, we also checked whether the use of
different sets of indices produces any significant discrepancy in the
estimated SSP parameters. We find that the estimated SSP
parameters agree well within the errors, although the use of
a larger set of indices naturally leads to smaller errors in the SSP
parameters. Therefore, we keep using all nine indices for the dE
nuclei in the further analysis.

\begin{figure}[h]
 \includegraphics[width=8cm]{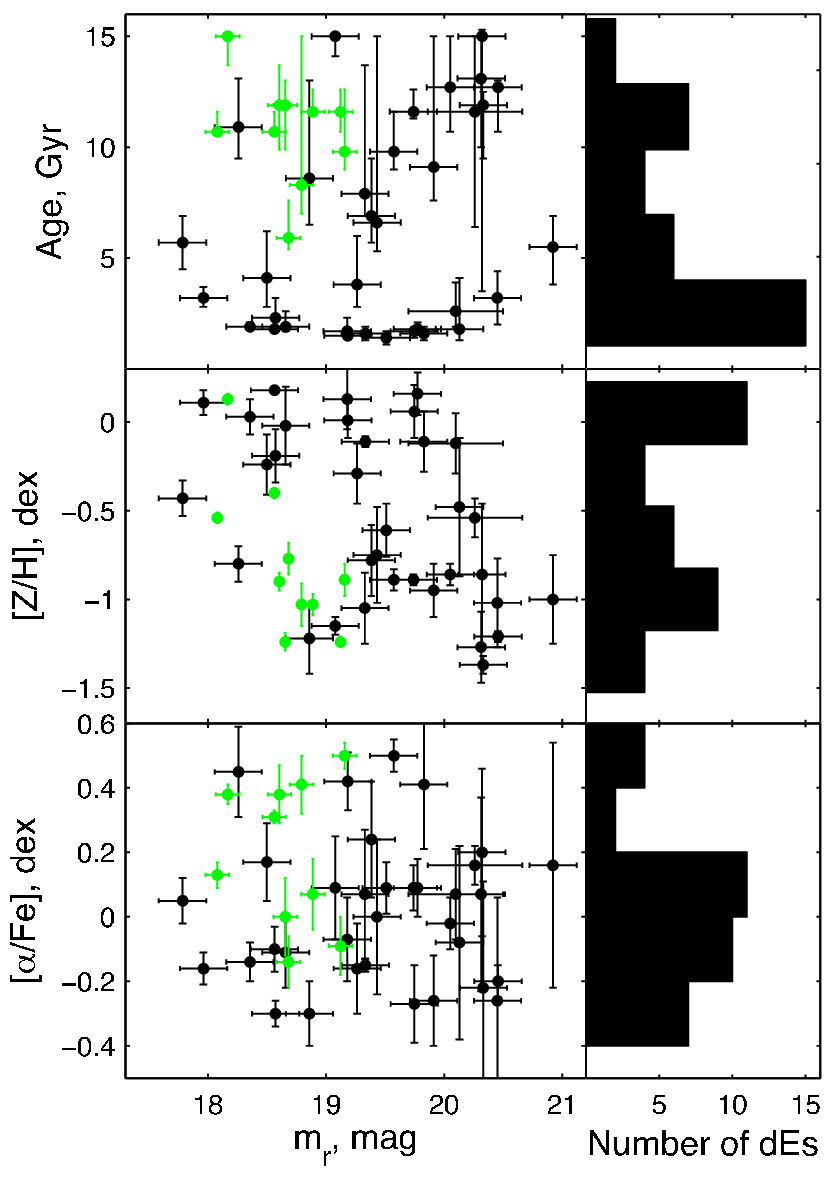}
 \caption{Age, metallicity and [$\alpha$/Fe]-abundance versus
   nucleus/UCD magnitude, for dE nuclei (black) and UCDs (green).}
 \label{env}
\end{figure}

The relation between the stellar population parameters and the local
projected galaxy density is plotted in Fig.~\ref{env2}. The local projected
density has been calculated from a circular projected area enclosing
the 10th neighbor. It seems that there is a weak correlation between
the local projected density and the ages of the nuclei. But, more
prominent than the correlation, an age break is seen at a projected
density $\approx$40 sq.deg.$^{-1}$. We thus use this value to divide our dE sample in
two groups, located in cluster regions of lower/higher density. We
find median ages of 2.6 Gyr for the low density group and 11.6 Gyr at
high densities.

Almost all UCDs lie in the high density region as defined
above. Their ages have a median of 11.7 Gyr, strikingly consistent
with the nuclei in the same density regime. As a statistical
comparison, we performed a Kolmogorov-Smirov (K-S) test with the null
hypothesis that there is no difference in the distribution of SSP
parameters between the UCDs and the dE nuclei of the high density
group. We find that there is a probability of 88\%, 97\% and 73\% for having the same
age, metallicity, and [$\alpha$/Fe]-abundance distribution, respectively.

The metallicities of dE nuclei also show a difference between the low
and high density regions: the median values differ by more than 0.5
dex, although the scatter is fairly large. The UCDs also have
a fairly large scatter in metallicity, ranging from +0.1 to $-$1.4
dex. However, it is remarkable that only two UCDs (VUCD3 and 4) have
a metallicity above $-$0.5 dex. The [$\alpha$/Fe]-abundances are
 more or less consistent with solar in case of the dE nuclei,
 but also a slight increase with density can be recognized
 for the nuclei and (more clearly) for the UCDs. Although the scatter is fairly
 large, the Spearman's rank correlation coefficients are 0.3 and 0.7 for dE nuclei and
 UCDs, with the probabilities for the null hypothesis that there is no
 correlation being 8.5\% and 1.5\%, respectively.
 In case of the UCDs, 7 out 10 have super-solar values of  [$\alpha$/Fe].

The relation between the stellar population
parameters and the $r$-band magnitudes of UCDs and nuclei is presented in
Fig.~\ref{env}. We do not see a relation of age
and [$\alpha$/Fe]-abundance with luminosity, both for nuclei and
UCDs. Unlike that, the metallicity of both UCDs and dE nuclei
  tends to increase with increasing luminosity, 
following the well known metallicity-luminosity relation. However,
bimodal peaks of the age and metallicity distributions of dE nuclei
can be seen as a striking feature in Fig~\ref{env} (right panel).
This indicates the inhomogeneous nature of dE nuclei, which can be divided into two different groups of stellar
population characteristics: a young metal-rich group and an old
metal-poor group. Brighter nuclei predominantly have metal-rich stellar
populations consistent with solar values. The metal-poor nuclei (metallicity $<$-0.5 dex) 
have an average of $-$1.0 dex, and most of them are less luminous than
$m_r=19.2$ mag ($M_r=-11.9$ mag).

\section{Summary and Discussion}
We have derived stellar population parameters of Virgo cluster UCDs and
nuclei of early-type dwarf galaxies, through fitting measured Lick
absorption line strengths to models of
\citet{Thomas03}. Since we use simple stellar population models,
it must be noted that our estimated parameters are interpreted as {\it
  SSP-equivalent.}

Indications that the UCDs in the Virgo cluster more likely have formed
via dE destruction in dense cluster regions than in the Fornax cluster
have already been presented in the literature \citep{Mieske06,
  Cote04acs, Cote06}. An explanation could be the higher mass of the
Virgo cluster, which provides a deeper potential well with stronger
tidal forces than in the Fornax cluster. Virgo UCDs were also found to
be significantly different from GCs in colors and sizes, as they
are bluer and larger than GCs \citep{Hasegan05}. E07 showed that Virgo
UCDs have structural properties similar to the Fornax
UCDs. Furthermore, \citet{Propris05} showed with the {\it HST} that Fornax UCDs are more extended and have higher
surface brightnesses than typical dwarf nuclei. That is not
necessarily inconsistent with the stripping scenarios, because the
stripping process can alter structural parameters of the embedded
nuclei. Therefore, a direct comparison of structural parameters may
not be able to constrain the formation scenarios of UCDs properly.

In the light of our new age and metallicity estimates in the previous
section, we want to address the possibility of formation of Virgo UCDs via
tidal destruction of nucleated dEs. We observe that UCDs exhibit significantly older and
more metal deficient stellar populations as compared to dE nuclei
\emph{on average}. However, if we compare the ages of
UCDs and nuclei only in high density regions, where most of the UCDs
are located, we do not find any significant difference in their age
and metallicity distribution. Only two dE nuclei (VCC0965 and VCC1122)
out of 13 from the high density
group have ages less than 5 Gyr and
metallicities above -0.2 dex. Note,
however, that due to projection effects, a number of objects that are
actually located in the outskirts of the cluster must fall into
regions of high \emph{projected} density, lying in front or
behind the center along the line of sight. If we consider the
low density group of our sample to be representative for nuclei of dEs
outside of the (threedimensional) cluster center, then we actually
\emph{expect} to have a few objects with young ages and high
metallicities in the high density group, consistent with what we
observe. Moreover, even without  taking this effect into account, our
statistical tests already show that the age distributions of UCDs and dE nuclei are
similar when considering only the group of nuclei in the high density.
This supports the idea of the threshing of dEs in high density environments
 to form the UCDs.

 At low densities, a certain fraction of dEs probably still retained some central
 gas and transformed it into stars (some dEs are still doing this, see
 \citealt{Lisker06b}). Consequently, their nuclei today would indeed show a younger stellar
population than a UCD, the latter being an ``old nucleus''. For such dEs, not only the
orbital average should be located in the outskirts, but they should also
not have passed right through the center (e.g.\ on an eccentric orbit),
otherwise the gas would have been stripped by ram pressure and tidal forces. This
would explain the fact that we do not see young UCDs: a dE destruction
is only possible if the galaxy really goes through the center and
experiences the strongest tidal forces. If we assume that 
the \emph{full}
destruction, i.e.\ the complete removal of the dE's main body, takes
significantly more orbital time than the gas stripping, then
nucleated dEs with an orbit leading through the center first lost
their gas, halting any star formation, and then became destroyed. Therefore,
we have no young UCDs.

In any case, young and bright dE nuclei
have either late or prolonged star formation activity. We assess
  this by modeling spectra of
  composite stellar populations through superposition of a young and
  an old component, using the model spectra of \cite{Vazdekis10}. We
fix the old population at an age of 11 Gyr and a metallicity of -1.2
dex, and add a component with solar metallicity and exponentially
declining star formation rate, starting from an age of 11 Gyr.
We find that spectra like those of the young nuclei (with an
SSP-equivalent age of
$\sim$3 Gyr) can be achieved when the exponentially declining
component formed stars until 2 Gyr ago, at which point it is
truncated. This means, if we assume our model to represent the
observed young nuclei, their star formation activity would have
stopped 9 Gyr later than that of the old nuclei.

On the one hand, the estimated ages and metallicities of UCDs nicely
agree with the ages and metallicities of nuclei of dEs from the dense
cluster regions. On the other hand, most of the old nuclei are fainter
than the UCDs. However, the UCD discoveries might suffer a selection
effect: if a \emph{faint} nucleus was stripped, it would now be
automatically counted into the globular cluster system of M87. We can
hardly estimate how many nucleated dEs have already been destroyed and
transformed into UCDs during the lifetime of the cluster. Thus we
cannot rule out or confirm any scenario \emph{just by counting} the
number of bright nuclei or that of UCDs. Only detailed future
simulations can resolve this issue.

 \section{Acknowledgments}

T.L.\ thanks Anna Pasquali and Eva Grebel for fruitful
discussions concerning UCDs, and Katharina Glatt for support
with the 2007 observing run. The authors thank G.\ Hensler, S.\ Kim,
R.\ Kotulla, H.\ Kuntschner, C.\ Mastropietro, S.-C.\ Rey, and
S.\ Weinmann for their valuable input to the 2009/10 observing
proposal and the data analysis.
   S.P.\ and T.L.\ are supported within the framework of the Excellence
    Initiative by the German Research Foundation (DFG) through the Heidelberg
    Graduate School of Fundamental Physics (grant number GSC 129/1). J.J. acknowledges support by the Gottlieb Daimler and Karl Benz Foundation. This study is based on the Sloan Digital Sky Survey
    (http://www.sdss.org).

\bibliographystyle{apj}

\end{document}